\documentclass{jpsj-suppl}
\usepackage{txfonts} 
\usepackage{wrapfig}

\title{How Well Do We Know The Supernova Equation of State?}

\author{Matthias \textsc{Hempel}$^{1}$, Micaela \textsc{Oertel}$^2$, Stefan \textsc{Typel}$^3$, and Thomas \textsc{Kl\"ahn}$^4$}

\inst{$^{1}$Department of Physics, University of Basel, Klingelbergstr. 82, CH-4056 Basel, Switzerland \\
$^{2}$LUTH, CNRS, Observatoire de Paris, Universit\'e Paris Diderot, 5 place Jules Janssen, F-92195 Meudon, France\\
$^{3}$GSI Helmholtzzentrum f\"ur Schwerionenforschung, Planckstra{\ss}e 1, D-64291 Darmstadt, Germany\\
$^{4}$Instytut Fizyki Teoretycznej, Uniwersytet Wroc{\l}awski, pl.~M.~Borna~9,  PL-50-204 Wroc{\l}aw, Poland}
\email{matthias.hempel@unibas.ch}

\recdate{August 20, 2016}

\abst{We give an overview about equations of state (EOS) which are currently available for simulations of core-collapse supernovae and neutron star mergers. A few selected important aspects of the EOS, such as the symmetry energy, the maximum mass of neutron stars, and cluster formation, are confronted with constraints from experiments and astrophysical observations. 
There are just very few models which are compatible even with this very restricted set of constraints. These remaining models illustrate the uncertainty of the uniform nuclear matter EOS at high densities. In addition, at finite temperatures the medium modifications of nuclear clusters represent a conceptual challenge. In conclusion, there has been significant development in the recent years, but there is still need for further improved general purpose EOS tables.}

\kword{equation of state, core-collapse supernova, neutron star, symmetry energy}

\begin{document}
\maketitle

\section{Introduction}

In simulations of core-collapse supernovae (CCSNe) the equation of state (EOS) has to cover
various different regimes: the progenitor star at the onset of collapse,
the formation of the proto-neutron star (PNS), and the development of the explosion
in the layers on top. The EOS has to be applicable at all these different 
evolutionary stages and spatial regions. 
It also should give a realistic description
of the remaining cold NS. Such an EOS is equally 
applicable in simulations of NS mergers, where 
various different densities, temperatures, and electron fractions (or equivalently 
isospin asymmetries) are encountered, similarly as in a CCSN. 
A better name for the ``supernova EOS'' would thus be ``general purpose EOS''. 

CCSNe and NS mergers are usually modeled using sophisticated hydrodynamic simulations.
Besides the neutrino-matter interactions, it is the EOS which provides the 
nuclear physics input. On the one hand, this happens via thermodynamic
quantities such as pressure or energy density, on the other hand via the 
nuclear composition. The latter information is required for the neutrino 
interactions. 

For cold NSs which are in beta-equilibrium without neutrinos, 
hundreds of different EOSs exist in the literature. Conversely,
just few general purpose EOS are available, 
due to the huge parameter range of
temperatures $0 \leq T \leq 100$~MeV, densities $10^4~\rm{g/cm^3} \leq \rho \leq 10^{15}~\rm{g/cm^3}$, and 
electron fractions $0 \leq Y_e \leq 0.6$,
which has to be covered. 
General purpose EOSs are usually provided and used in tabular form.
There are three crucial aspects of the general purpose EOS, which will be addressed
briefly in the following: (1) some properties of uniform nuclear matter, (2) the 
formation of nuclei at low densities, and (3) additional non-nucleonic degrees
of freedom at high densities and/or temperatures.

\section{Properties of Uniform Nuclear Matter}

\begin{table}
\caption{Based on \cite{eos_rmp}. Currently existing general purpose EOSs. Listed are the
  nuclear interaction model used, the maximum mass $M_{\rm max}$ of cold
  NSs, and corresponding references.
  \label{tab:eos3d}}
\begin{minipage}[t]{.5\linewidth}
\centering
EOSs containing nucleons and nuclei.
\begin{tabular}{c| c c l}
\hline 
\hline 
Model              & Nuclear     & $M_{\rm max}$ & References \\ 
                   & Interaction & (M$_\odot$)   & \\ 
\hline
H\&W               & SKa      & 2.21 &\cite{eid80,hillebrandt84} \\ 
LS180              & LS180    & 1.84 &\cite{lattimer91} \\ 
LS220              & LS220    & 2.06 &\cite{lattimer91} \\ 
LS375              & LS375    & 2.72 &\cite{lattimer91} \\ 
STOS               & TM1      & 2.23 &\cite{shen98,shen98_2,shen11} \\ 
FYSS               & TM1      & 2.22 &\cite{furusawa13_eos} \\ 
HS(TM1)            & TM1      & 2.21 &\cite{hempel10,hempel11} \\ 
HS(TMA)            & TMA      & 2.02 &\cite{hempel10} \\ 
HS(FSU)            & FSUgold  & 1.74 &\cite{hempel10,hempel11} \\ 
HS(NL3)            & NL3      & 2.79 &\cite{hempel10,fischer14} \\ 
HS(DD2)            & DD2      & 2.42 &\cite{hempel10,fischer14} \\ 
HS(IUFSU)          & IUFSU    & 1.95 &\cite{hempel10,fischer14} \\ 
SFHo               & SFHo     & 2.06 &\cite{steiner13} \\ 
SFHx               & SFHx     & 2.13 &\cite{steiner13} \\ 
SHT(NL3)           & NL3      & 2.78 &\cite{shen2011a} \\ 
SHO(FSU)           & FSUgold  & 1.75 &\cite{shen2011b} \\ 
SHO(FSU2.1)        & FSUgold2.1&2.12 &\cite{shen2011b} \\ 
\hline 
\hline
\end{tabular}
\end{minipage}
\hfill
\begin{minipage}[t]{.5\linewidth}
\centering
EOSs including additional degrees of freedom.
\begin{tabular}{c| c c l}
\hline 
\hline 
Model              & Nuclear     & $M_{\rm max}$ & References \\ 
                   & Interaction & (M$_\odot$)   & \\ 
\hline
LS220$\Lambda$     & LS220    & 1.91 &\cite{Oertel:2012qd,Gulminelli:2013qr} \\ 
LS220$\pi$         & LS220    & 1.95 &\cite{Oertel:2012qd,Peres_13} \\ 
BHB$\Lambda$       & DD2      & 1.96 &\cite{banik14} \\ 
BHB$\Lambda\phi$   & DD2      & 2.11 &\cite{banik14} \\ 
STOS$\Lambda$      & TM1      & 1.90 &\cite{shen11} \\ 
STOSYA30           & TM1      & 1.59 &\cite{ishizuka08} \\ 
STOSYA30$\pi$      & TM1      & 1.62 &\cite{ishizuka08} \\ 
STOSY0             & TM1      & 1.64 &\cite{ishizuka08} \\ 
STOSY0$\pi$        & TM1      & 1.67 &\cite{ishizuka08} \\ 
STOSY30            & TM1      & 1.65 &\cite{ishizuka08} \\ 
STOSY30$\pi$       & TM1      & 1.67 &\cite{ishizuka08} \\ 
STOSY90            & TM1      & 1.65 &\cite{ishizuka08} \\ 
STOSY90$\pi$       & TM1      & 1.67 &\cite{ishizuka08} \\ 
STOS$\pi$          & TM1      & 2.06 &\cite{nakazato08} \\ 
STOSQ209n$\pi$     & TM1      & 1.85 &\cite{nakazato08} \\ 
STOSQ162n          & TM1      & 1.54 &\cite{nakazato13} \\ 
STOSQ184n          & TM1      & 1.36 &\cite{nakazato13} \\ 
STOSQ209n          & TM1      & 1.81 &\cite{nakazato08,nakazato13} \\ 
STOSQ139s          & TM1      & 2.08 &\cite{fischer14b} \\ 
STOSQ145s          & TM1      & 2.01 &\cite{sagert12} \\ 
STOSQ155s          & TM1      & 1.70 &\cite{fischer11} \\ 
STOSQ162s          & TM1      & 1.57 &\cite{sagert09} \\ 
STOSQ165s          & TM1      & 1.51 &\cite{sagert09} \\ 
\hline 
\hline
\end{tabular}
\end{minipage}
\end{table}

At present, there are 17 general purpose EOS tables available which consider
nucleons and nuclei as particles degrees of freedom, as listed in the left part 
of Table~\ref{tab:eos3d}. There 
are much more which are still in preparation, and other, equally important 
works about the EOS in NS (mergers) and CCSNe, which just concentrate on parts
of the parameter space and do not aim at providing a full table. All of the 
17 available ``nucleonic'' general purpose EOSs employ mean-field interactions
of the nucleons and a phenomenological description of nuclei. Thus it is important
to confront these models with experimental data and more solid 
theoretical nuclear matter calculations.
Some of the 17 EOS tables share the same nucleon interactions. In total,
only 13 different interactions are used, as can be seen in Table~\ref{tab:eos3d}. 
Here we neglect small differences, such as the usage of different nucleon masses
or minor changes at low densities, as employed, e.g., in \cite{shen10,hempel11}.
For further details, see \cite{eos_rmp}.

\subsection{Symmetry Energy}
\label{esym}
The symmetry energy describes how matter behaves when going to neutron-rich
conditions, and thus is quite important for NSs and CCSNe. Many experimental
constraints for the symmetry energy are available in the literature. 
In a careful analysis of several experimental probes, Lattimer and Lim \cite{lattimer2013} derive 
$J=(29.0-32.7)$~MeV and $L=(40.5-61.9)$~MeV for the value of the symmetry 
energy $J$ and the slope parameter $L$, both at saturation density $n_B^0$.
Using this constraint, only the general purpose EOSs based on 
IUFSU, SFHo, DD2, FSUgold, and FSUgold2.1 remain as viable models
as can be seen in Fig.~\ref{fig:jl}. Note that FSUgold2.1 has the same values of $J$ and $L$ as
FSUgold, and the values of the three LS versions are all the same.

\subsection{Neutron Star Maximum Mass}
The recent observations of two NSs with masses around 2~M$_\odot$ 
\cite{antoniadis2013,demorest2010,fonseca16} put
an important constraint on the high-density part of the EOS. Four of the
17 nucleonic general purpose EOSs are not able to reach the lower 1-sigma limit 
of \cite{antoniadis2013} of 1.97~M$_\odot$. These are LS180, HS(IUFSU), HS(FSU), and 
SHO(FSU). The maximum mass of HS(IUFSU) with 1.95~M$_\odot$ is just slightly too low.
The last two models are both based on the FSUgold interactions, 
and thus have a very similar mass-radius curve and a low maximum mass around 
1.74~M$_\odot$. Note that the FSUgold2.1 interaction used in SHO(FSU2.1) is a modification
of FSUgold, where an additional pressure contribution has been added at high 
densities to increase the maximum mass sufficiently.


\subsection{Viable Models}
Even if we use just the two constraints for the symmetry energy 
($J$ and $L$), and the maximum mass, most of the general purpose EOSs
are ruled out. Only three of 17 pass these two tests: HS(DD2), SFHo, and SHO(FSU2.1).
One always has to be careful with such a simple pass-fail classification. 
Obviously, the relevance of any constraint depends on the 
context where the EOS is applied. For example in a 
CCSN of a very light progenitor, the mass of the PNS stays much below
2~M$_\odot$. Thus for this particular scenario 
even an EOS with a too low maximum mass might still be acceptable. On the 
other hand, it is clear that such an EOS would not be the most realistic one
for a general context. 
\begin{wrapfigure}{L}{0.35\textwidth}
\includegraphics{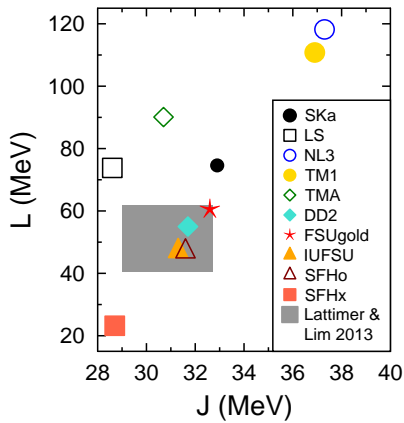}
\caption{Symmetry energy at saturation density $J$ and slope parameter $L$ for the
interactions of the general purpose EOSs in comparison with the constraint of \cite{lattimer2013}.}
\label{fig:jl}
\end{wrapfigure}

Here we have chosen only the constraints for the symmetry energy and the
maximum mass, because they belong to the most important for the 
physics of compact stars and they are well studied and are among the most robust.
Of course there are much more constraints available, see also \cite{dutra2012,dutra2014}. 
If one would apply all existing constraints blindly, probably all EOSs would 
be excluded, as different constraints sometimes contradict each
other. But even considering just a few more constraints could be
problematic. For example DD2 does not fulfill the so-called flow-constraint 
of \cite{danielewicz02}, and SFHo shows some minor deviations for the neutron 
matter EOS compared to theoretical constraints from the Chiral EFT calculations of
\cite{tews13}, see \cite{fischer14}.

\section{Description of Nuclei}
The different EOS models make different simplifications regarding the 
nuclear degrees of freedom at low densities. For light nuclei, often only the alpha particle
is considered (H\&W, LS, STOS, SHT(NL3), SHO(FSU), SHO(FSU2.1)). 
For heavy nuclei, in some models the single nucleus approximation 
is used (LS, STOS), where the thermal distribution of different nuclear species
is replaced by a single representative heavy nucleus. 

\subsection{Light Nuclei}
It has been found in a number of works that light clusters, such as deuterons
or tritons, can be more abundant than free protons in the shock-heated matter of
a CCSN, see, e.g., \cite{sumiyoshi08,hempel11,fischer14}. High abundances of 
light nuclei are found in the vicinity of the neutrino spheres, and thus
one can expect that they might have an impact on the neutrino luminosities and
mean energies. It has been found that
they can modify the neutrino-driven wind \cite{oconnor07,arcones08}, which is
important for nucleosynthesis. Their effect on the 
supernova dynamics and neutrino quantities is still not completely settled, 
but probably it is only moderate, see 
\cite{furusawa2013,fischer16}.

Nuclear clusters are also formed in heavy-ion collisions, under very similar
thermodynamic conditions as in a CCSN. Qin et al.\ 
\cite{qin12} measured charged particle and neutron yields at Texas A\&M and also extracted 
temperatures and densities at different stages of the collision from this data. 
They compared the experimental data with predictions of some of the general purpose EOSs. However,
they did not take into account the differences between matter in the experiment and
the astrophysical environment, such as charge neutrality and Coulomb interactions
in supernova matter or the limited number of nucleons involved in the heavy-ion
collision. 

Using the same experimental data of \cite{qin12}, these aspects were considered in \cite{hempel15} 
Also some errors
in the theoretical EOS data shown in \cite{qin12} were corrected. 
From the comparison it was concluded
that the ideal gas behavior is clearly ruled out at similar densities as they occur 
around the neutrino-spheres in a CCSN. To obtain a good
agreement with the experimental data, three ingredients seem to be necessary:
(i) inclusion of all relevant particle degrees of freedom, (2) mean-field 
interactions of unbound nucleons, (3) a suppression mechanism of nuclei at 
high densities (e.g., Pauli-blocking or excluded volume). The largest
deviations were found for the LS EOSs which show a notable underproduction
of alpha particles and/or too many nucleons, and for the SHT(NL3) and SHO(FSU2.1) EOSs, with too many
alpha particles and/or too little nucleons predicted. The current experimental
data is not accurate enough to distinguish details of the medium modifications 
of the light clusters.

It has to be noted that this experimental constraint does not have
the same significance and robustness as, e.g., the study of \cite{lattimer2013} or the 
observed pulsar masses, as some model assumptions are involved in the extraction
of the data.
Nevertheless, the obtained conclusions appear to be 
reasonable. Found deviations can be linked conclusively to deficits or missing aspects
in the theoretical models, and for the most advanced models regarding 
medium modifications of light clusters \cite{typel09,roepke2014b}, good agreement with the experimental 
data is found. 

\subsection{Heavy Nuclei}
Heavy nuclei dominate the composition in a CCSN during the early stage of the 
collapse of the iron core. Later, they are present in the matter which is 
accreted onto the PNS from the outer layers of the progenitor. 
They are especially relevant for the electron-captures during the collapse,
which affects, e.g., the mass of the core at bounce. In \cite{buyuk13},
three statistical models, HS, FYSS, and SMSM \cite{botvina08},
 were compared, which all go beyond the single nucleus
approximation for heavy nuclei, i.e., which contain an ensemble of various
different species. On the one hand, overall similar trends
are found for various densities, temperatures, and asymmetries. On the other 
hand, the limitations and simplifications used in each of the models
are visible in the details of the nuclear composition,
especially when going to more extreme conditions.

\section{Additional Degrees of Freedom}
At high densities or temperatures additional degrees of freedom such as pions, hyperons,
or quarks can appear. Currently 23 general purpose EOSs exist which consider
such additional degrees of freedom. They are listed in the right half of Table~\ref{tab:eos3d}.
Most of them employ the TM1 interactions for the nucleons, whose symmetry energy is problematic
as the values of $J$ and $L$ are much higher than in experimental constraints,
see Fig.~\ref{fig:jl}.
Models which do not employ TM1 are 
LS220$\Lambda$ and LS220$\pi$, representing extensions of LS220,
and BHB$\Lambda$ and BHB$\Lambda \phi$, representing extensions of HS(DD2).
As the symmetry energy of LS220 is also slightly too high in comparison with the constraints 
from \cite{lattimer2013}, only BHB$\Lambda$ and BHB$\Lambda \phi$ remain as 
directly compatible models. 

Additional degrees of freedom often lead to a substantial reduction of the maximum
mass. Even if one uses only the maximum mass constraint, just few EOSs remain as viable models:
only STOSQ139s, STOSQ145s, STOS$\pi$, and BHB$\Lambda \phi$ are in direct agreement with the constraint from \cite{antoniadis2013}. The maximum masses of BHB$\Lambda$ and LS220$\pi$
are slightly too low. All others have maximum masses around or below 1.9~M$_\odot$.
Using the two constraints for the symmetry energy and the maximum mass together,
BHB$\Lambda \phi$ is the only general purpose EOS which fulfills both of them. 
In this model only the Lambda of all possible hyperons has been added to 
the HS(DD2) EOS. 

\section{Summary and Conclusions}
The supernova or general purpose EOS has to 
cover a huge parameter space in density, temperature, and electron fraction.
It remains a conceptual and numerical challenge to develop an EOS model and 
to calculate an EOS table covering all the different thermodynamic regimes. 
As a consequence, only a very limited number of general purpose EOSs is 
currently available. All of them are to a large extent based on a
phenomenological description.
Advances in nuclear experiments, astrophysical observations,
but also in theoretical ab-initio calculations allow to significantly constrain 
the general purpose EOS. Here we discussed constraints for the symmetry energy
and its slope parameter, the neutron star maximum mass, and cluster yields from 
heavy-ion collisions. 

Regarding the nuclear composition, even though the models give 
overall similar trends for varying temperature, density and asymmetry, there is still
some significant model dependency remaining. 
Further improvements of the general purpose EOS regarding the description of heavy and light nuclei
and their medium modifications, within a statistical distribution, is clearly
demanded. One of the most advanced models in this respect 
is FYSS, which implements aspects of the more systematic and microscopic approaches of
\cite{typel09,roepke2014b}, but also in this EOS still a lot of phenomenological 
modeling is involved.

Even if one considers only the maximum mass and the symmetry energy constraint,
just three nucleonic general purpose EOS remain as viable models: 
HS(DD2), SFHo, and SHO(FSU2.1). These remaining three models can be taken as 
a representative sample illustrating the current uncertainties of the nucleon
interactions at high densities. It has to be noted that all of them 
have aspects which can be improved further.

Regarding general purpose EOSs which address the appearance of additional 
degrees of freedom at high densities and/or temperatures (pions, hyperons, 
and/or quarks) the situation is even more severe. Just four EOSs exist 
which have a sufficiently high maximum mass.
Only one EOS, BHB$\Lambda \phi$, an extension of HS(DD2) where
lambda hyperons have been added, passes both constraints. 
This means at present there is no general purpose EOS which
has a good behavior of the symmetry energy, a maximum mass above 2~M$_\odot$,
and which considers pions or quarks. This situation is certainly 
not satisfactory and should be improved.
From our perspective it is not justified to simply ignore these additional 
degrees of freedom. New, further improved general purpose EOSs should be 
developed in the future, also exploring these additional degrees of freedom.

\section*{Acknowledgement}
This work has been partially funded by the SN2NS project ANR-10-BLAN-0503 and by
NewCompStar, COST Action MP1304. 
M.H.\ is supported by the Swiss National Science Foundation. 
T.K.\ is grateful for support by the Polish National
Science Center (NCN) under grant number UMO-2013/09/B/ST2/01560.
S.T.\ is supported by the Helmholtz Association (HGF) through the
Nuclear Astrophysics Virtual Institute (VH-VI-417).


\begin{thebibliography}{9}
\bibitem{eos_rmp}
M. Oertel, M. Hempel, T. Kl\"ahn, and S. Typel:  submitted to Rev. Mod. Phys.
  (2016) .
\bibitem{eid80}
M.~F. {El Eid} and W. {Hillebrandt}:  {Astron. Astrophys. Suppl. Ser.} {\bf 42}
  (1980) 215.
\bibitem{hillebrandt84}
W. {Hillebrandt}, K. {Nomoto}, and R.~G. {Wolff}:  Astron. Astrophys. {\bf 133}
  (1984) 175.
\bibitem{lattimer91}
J.~M. {Lattimer} and D.~F. {Swesty}:  Nucl. Phys. A {\bf 535} (1991) 331.
\bibitem{shen98}
H. {Shen}, H. {Toki}, K. {Oyamatsu}, and K. {Sumiyoshi}:  Nucl. Phys. A {\bf
  637} (1998) 435.
\bibitem{shen98_2}
H. {Shen}, H. {Toki}, K. {Oyamatsu}, and K. {Sumiyoshi}:  Prog. Theor. Phys.
  {\bf 100} (1998) 1013.
\bibitem{shen11}
H. {Shen}, H. {Toki}, K. {Oyamatsu}, and K. {Sumiyoshi}:  Astrophys. J. Suppl.
  {\bf 197} (2011) 20.
\bibitem{furusawa13_eos}
S. {Furusawa}, K. {Sumiyoshi}, S. {Yamada}, and H. {Suzuki}:  Astrophys. J.
  {\bf 772} (2013) 95.
\bibitem{hempel10}
M. {Hempel} and J. {Schaffner-Bielich}:  Nucl. Phys. A {\bf 837} (2010) 210.
\bibitem{hempel11}
M. {Hempel}, T. {Fischer}, J. {Schaffner-Bielich}, and M. {Liebend{\"o}rfer}:
  Astrophys. J. {\bf 748} (2012) 70.
\bibitem{fischer14}
T. {Fischer}, M. {Hempel}, I. {Sagert}, Y. {Suwa}, and J. {Schaffner-Bielich}:
  Eur. Phys. J. A {\bf 50} (2014) 46.
\bibitem{steiner13}
A.~W. {Steiner}, M. {Hempel}, and T. {Fischer}:  Astrophys. J. {\bf 774} (2013)
  17.
\bibitem{shen2011a}
G. {Shen}, C.~J. {Horowitz}, and S. {Teige}:  Phys. Rev. C {\bf 83} (2011)
  035802.
\bibitem{shen2011b}
G. {Shen}, C.~J. {Horowitz}, and E. {O'Connor}:  Phys. Rev. C {\bf 83} (2011)
  065808.
\bibitem{Oertel:2012qd}
M. {Oertel}, A.~F. {Fantina}, and J. {Novak}:  Phys. Rev. C {\bf 85} (2012)
  055806.
\bibitem{Gulminelli:2013qr}
F. Gulminelli, A. Raduta, M. Oertel, and J. Margueron:  Phys. Rev. C {\bf 87}
  (2013) 055809.
\bibitem{Peres_13}
B. {Peres}, M. {Oertel}, and J. {Novak}:  Phys. Rev. D {\bf 87} (2013) 043006.
\bibitem{banik14}
S. {Banik}, M. {Hempel}, and D. {Bandyopadhyay}:  Astrophys. J. Suppl. {\bf
  214} (2014) 22.
\bibitem{ishizuka08}
C. {Ishizuka}, A. {Ohnishi}, K. {Tsubakihara}, K. {Sumiyoshi}, and S. {Yamada}:
   J. Phys. G Nucl. Phys. {\bf 35} (2008) 085201.
\bibitem{nakazato08}
K. {Nakazato}, K. {Sumiyoshi}, and S. {Yamada}:  Phys. Rev. D {\bf 77} (2008)
  103006.
\bibitem{nakazato13}
K. {Nakazato}, K. {Sumiyoshi}, and S. {Yamada}:  Astron. Astrophys. {\bf 558}
  (2013) A50.
\bibitem{fischer14b}
T. {Fischer}, T. {Kl{\"a}hn}, I. {Sagert}, M. {Hempel}, and D. {Blaschke}:
  Acta Phys. Polon. Supp. {\bf 7} (2014) 153.
\bibitem{sagert12}
I. {Sagert}, T. {Fischer}, M. {Hempel}, G. {Pagliara}, J. {Schaffner-Bielich},
et al.:
Acta Phys. Polon. B {\bf 43}
  (2012) 741.
\bibitem{fischer11}
T. {Fischer}, I. {Sagert}, G. {Pagliara}, M. {Hempel}, J. {Schaffner-Bielich},
et al.:
Astrophys. J. Suppl.
  {\bf 194} (2011) 39.
\bibitem{sagert09}
I. {Sagert}, M. {Hempel}, G. {Pagliara}, J. {Schaffner-Bielich}, T. {Fischer},
et al.:
Phys. Rev.
  Lett. {\bf 102} (2009) 081101.
\bibitem{shen10}
G. {Shen}, C.~J. {Horowitz}, and S. {Teige}:  Phys. Rev. C {\bf 82} (2010)
  015806.
\bibitem{lattimer2013}
J.~M. {Lattimer} and Y. {Lim}:  Astrophys. J. {\bf 771} (2013) 51.
\bibitem{antoniadis2013}
J. {Antoniadis}, P.~C.~C. {Freire}, N. {Wex}, T.~M. {Tauris}, R.~S. {Lynch},
et al.:
Science
  {\bf 340} (2013) 448.
\bibitem{demorest2010}
P.~B. {Demorest}, T. {Pennucci}, S.~M. {Ransom}, M.~S.~E. {Roberts}, and
  J.~W.~T. {Hessels}:  Nature {\bf 467} (2010) 1081.
\bibitem{fonseca16}
E. {Fonseca}, T.~T. {Pennucci}, J.~A. {Ellis}, I.~H. {Stairs}, D.~J. {Nice},
et al.:
  arXiv:1603.00545 (2016) .
\bibitem{dutra2012}
M. {Dutra}, O. {Louren{\c c}o}, J.~S. {S{\'a} Martins}, A. {Delfino}, J.~R.
  {Stone}, 
et al.:
Phys. Rev. C {\bf 85} (2012) 035201.
\bibitem{dutra2014}
M. {Dutra}, O. {Louren{\c c}o}, S.~S. {Avancini}, B.~V. {Carlson}, A.
  {Delfino}, 
et al.:
Phys. Rev. C {\bf 90} (2014) 055203.
\bibitem{danielewicz02}
P. {Danielewicz}, R. {Lacey}, and W.~G. {Lynch}:  Science {\bf 298} (2002)
  1592.
\bibitem{tews13}
I. {Tews}, T. {Kr{\"u}ger}, K. {Hebeler}, and A. {Schwenk}:  Phys. Rev. Lett.
  {\bf 110} (2013) 032504.
\bibitem{sumiyoshi08}
K. {Sumiyoshi} and G. {R{\"o}pke}:  Phys. Rev. C {\bf 77} (2008) 055804.
\bibitem{oconnor07}
E. {O'Connor}, D. {Gazit}, C.~J. {Horowitz}, A. {Schwenk}, and N. {Barnea}:
  Phys. Rev. C {\bf 75} (2007) 055803.
\bibitem{arcones08}
A. {Arcones}, G. {Mart{\'{\i}}nez-Pinedo}, E. {O'Connor}, A. {Schwenk}, H.-T.
  {Janka}, 
et al.:
Phys. Rev. C {\bf 78} (2008)
  015806.
\bibitem{furusawa2013}
S. {Furusawa}, H. {Nagakura}, K. {Sumiyoshi}, and S. {Yamada}:  Astrophys. J.
  {\bf 774} (2013) 78.
\bibitem{fischer16}
T. {Fischer}, G. {Mart{\'{\i}}nez-Pinedo}, M. {Hempel}, L. {Huther}, G.
  {R{\"o}pke}, 
et al.:
Eur. Phys. J. Web Conf. {\bf 109}
  (2016) 06002.
\bibitem{qin12}
L. {Qin}, K. {Hagel}, R. {Wada}, J.~B. {Natowitz}, S. {Shlomo}, 
et al.:
Phys. Rev. Lett. {\bf 108} (2012) 172701.
\bibitem{hempel15}
M. {Hempel}, K. {Hagel}, J. {Natowitz}, G. {R\"opke}, and S. {Typel}:  Phys.
  Rev. C {\bf 91} (2015) 045805.
\bibitem{typel09}
S. {Typel}, G. {R{\"o}pke}, T. {Kl{\"a}hn}, D. {Blaschke}, and H.~H. {Wolter}:
  Phys. Rev. C {\bf 81} (2010) 015803.
\bibitem{roepke2014b}
G. {R{\"o}pke}:  Phys. Rev. C {\bf 92} (2015) 054001.
\bibitem{buyuk13}
N. {Buyukcizmeci}, A.~S. {Botvina}, I.~N. {Mishustin}, R. {Ogul}, M. {Hempel},
et al.:
Nucl. Phys. A {\bf 907} (2013) 13.
\bibitem{botvina08}
A.~S. {Botvina} and I.~N. {Mishustin}:  Nucl. Phys. A {\bf 843} (2010) 98.
\end{thebibliography}

\end{document}